\documentclass[]{PoS}
\def\deg{\ifmmode^\circ\else$^\circ$\fi}

\title{Winds in Collision: high-energy particles in massive binary systems}

\ShortTitle{Massive binary systems}

\author{\speaker{Sean M. Dougherty} \thanks{This work has been done in collaboration with
Perry Williams (U. Edinburgh, UK), Sven Van Loo (U. Leeds, UK), Tony
Beasley (ALMA), Ronny Blomme (Royal Observatory, Belgium), Evan
O'Connor (UPEI, Canada) and Nick Bolingbroke (U. Victoria, Canada)}\\
National Research Council Herzberg Institute for Astrophysics,
Canada\\ E-mail: \email{sean.dougherty@nrc.ca}}

\author{Julian M. Pittard\\
        Physics and Astronomy, University of Leeds, UK\\
        E-mail: \email{jmp@ast.leeds.ac.uk}}

\abstract{High-resolution radio observations have revealed that
non-thermal radio emission in WR stars arises where the stellar wind
of the WR star collides with that of a binary companion.  These
colliding-wind binary (CWB) systems offer an important laboratory for
investigating the underlying physics of particle
acceleration. Hydrodynamic models of the binary stellar winds and the
wind-collision region (WCR) that account for the evolution of
the electron energy spectrum, largely due to inverse Compton cooling,
are now available. Radiometry and imaging obtained with the VLA,
MERLIN, EVN and VLBA provide essential constraints to these
models. Models of the radio emission from WR146 and WR147 are shown,
though these very wide systems do not have defined orbits and hence
lack a number of important model parameters.  Multi-epoch VLBI imaging
of the archetype WR+O star binary WR140 through a part of its 7.9-year
orbit has been used to define the orbit inclination, distance and
the luminosity of the companion star to enable the best constraints
for any radio emitting CWB system.  Models of the spatial distribution
of relativistic electrons and ions, and the magnetic energy density
are used to model the radio emission, and also to predict the high
energy emission at X-ray and $\gamma$-ray energies. It is clear that
high-energy facilities e.g. GLAST and VERITAS, will be important for
constraining particle acceleration parameters such as the spectral
index of the energy spectrum and the acceleration efficiency of both
ions and electrons, and in turn, identify unique models for the radio
spectra. This will be especially important in future attempts to model
the spectra of WR140 throughout its complete orbit. A WCR origin for
the synchrotron emission in O-stars, the progenitors of WR stars, is
illustrated by observations of Cyg OB2 \#9.}

\FullConference{The 8th European VLBI Network Symposium on New
                Developments in VLBI Science and Technology and EVN
                Users Meeting \\ September 26-29 2006\\ Torun, Poland}

\begin{document}

\section{Wind-collision regions and particle acceleration}
Wolf-Rayet (WR) stars have dense stellar winds that are photo-ionized
by the strong UV radiation fields from the underlying WR star, giving
rise to a free-free continuum emission spectra, observable from IR to
radio wavelengths.  The brightness temperature of this emission is
$\sim10^4$~K, as expected from a photo-ionized envelope in thermal
equilibrium. The emission is partially optically thick, and between
mid-IR and radio frequencies has a power-law spectrum
$S_\nu\propto\nu^\alpha$, with spectral index ($\alpha$) typically
$\sim+0.7$ e.g. \cite{Williams:1990}.  A number of WR stars have
radio emission properties that differ from this typical picture: they
exhibit high brightness temperatures ($\sim10^6-10^7$~K), and have
flat or negative spectral indices, properties characteristic of
non-thermal synchrotron emission and therefore of high-energy
phenomena in the stellar winds \cite{Abbott:1984, Abbott:1986}.

\begin{figure*}
\includegraphics[width=0.47\textwidth]{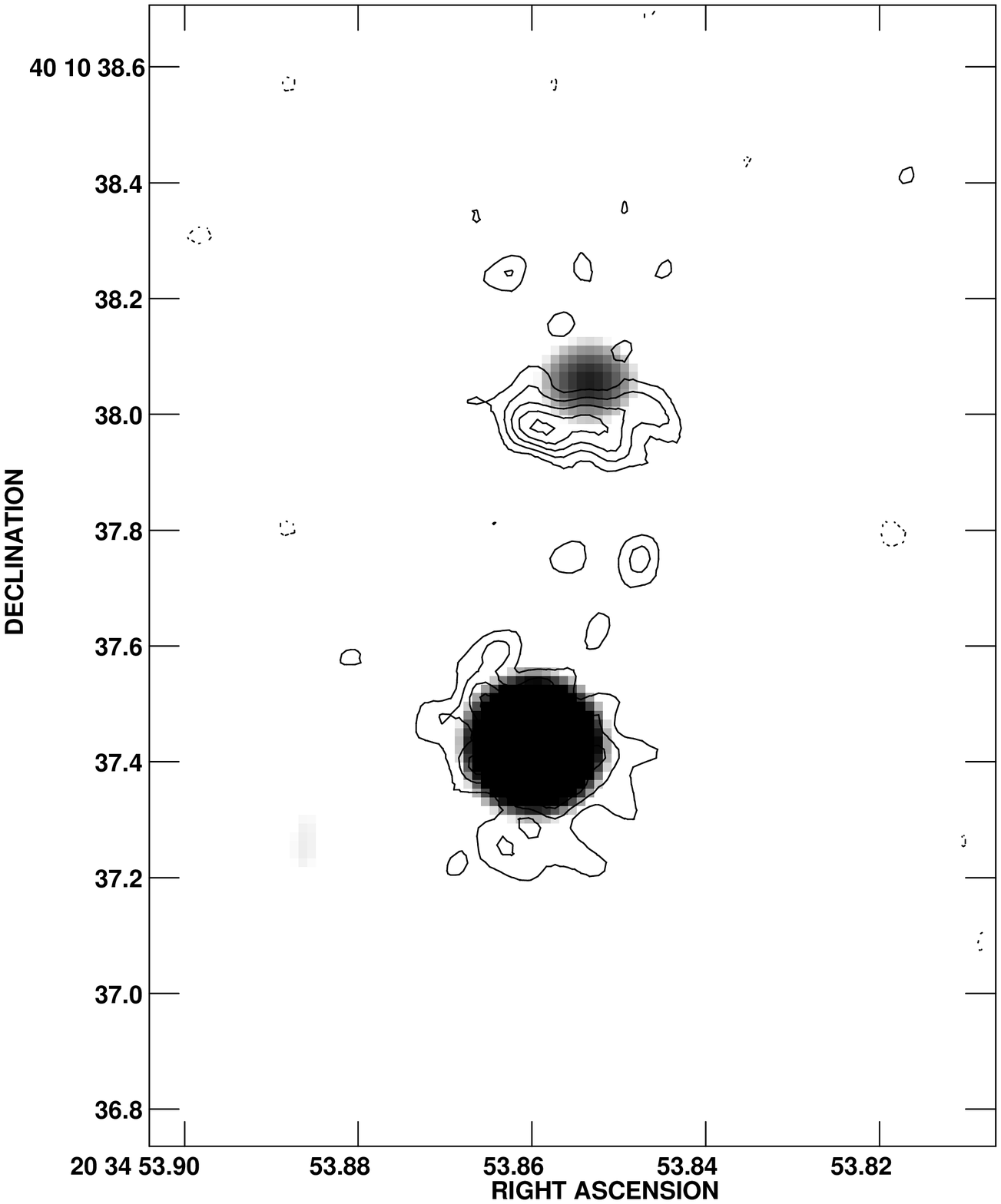}
\includegraphics[bb=28 85 567 630, width=0.54\textwidth]{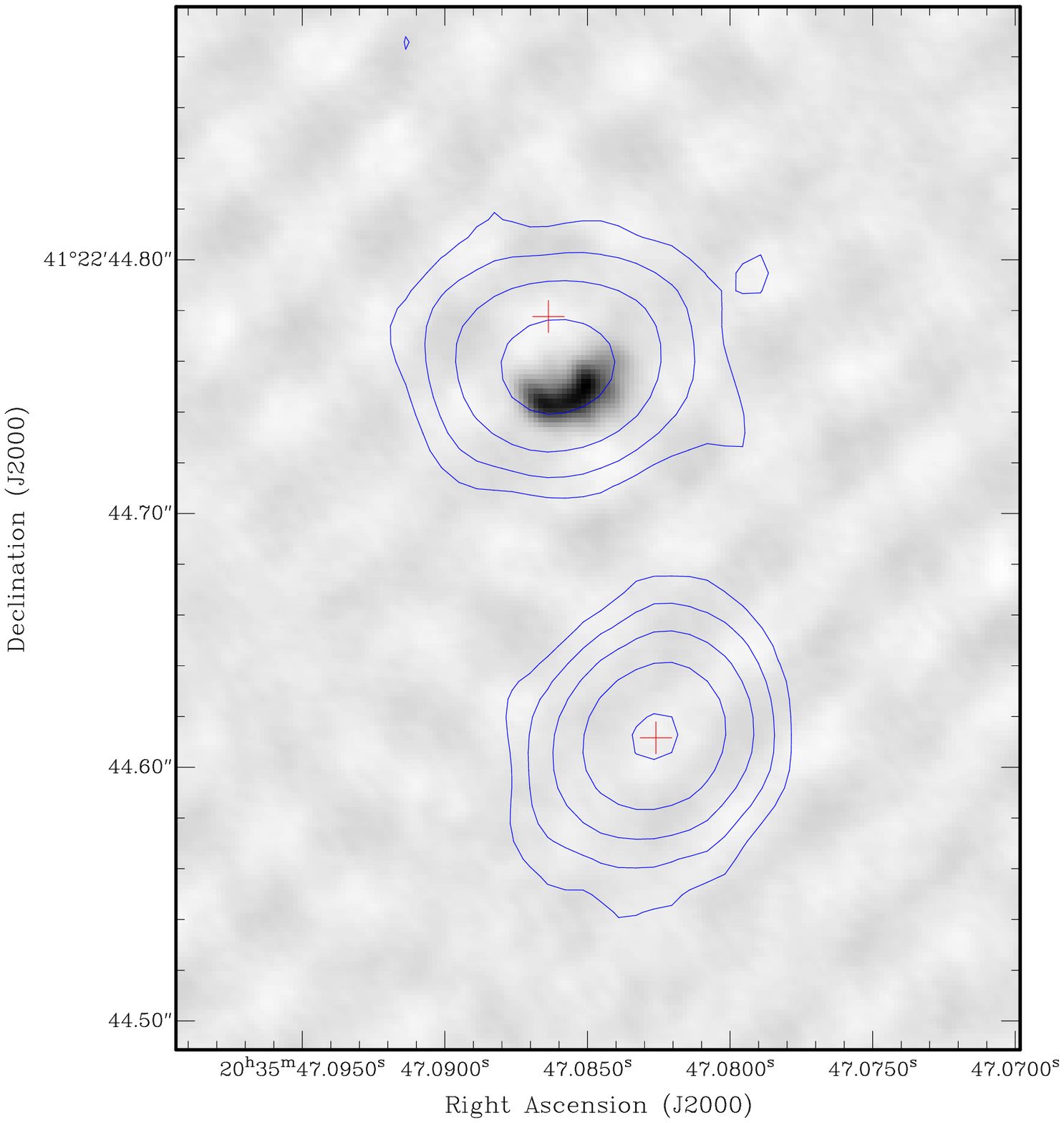}
\caption[]{Left: Overlay of a MERLIN 5-GHz image of WR147 (contours)
and a $2.2\mu$m UKIRT image (greyscale). The IR image of the WR star
was aligned with the peak position of the thermal emission in the
southern radio component, assumed to be the ionized stellar wind of
the WR star. The position of the non-thermal emission region (the
northern radio component) is consistent with the location of ram
pressure balance between the winds of the WR star and the B-type
companion revealed in the IR image \cite{Williams:1997}. Right:
4.8-GHz EVN (greyscale) and 43-GHz VLA+PT (contours) observations of
WR146. The crosses denote the relative positions of the stars
determined by HST \cite{Niemela:1998} and assuming the peak of the
southern 43-GHz component (thermal emission) is the WR star
(from \cite{O'Connor:2005}).  The northern component is dominated by
synchrotron emission, even at 43 GHz c.f. Fig.~\ref{fig:146}. }
\label{fig:147}
\end{figure*}

Spatially resolved observations of the WR+O binary systems WR~147
\cite{Niemela:1998, Williams:1997} and WR~146
\cite{Dougherty:2000a,Dougherty:1996} presented the first unequivocal
confirmation that the synchrotron emission did not arise within the
stellar wind of a single star, but at the location of the
wind-collision region (WCR) where the stellar winds of two massive
stars collide - the colliding-wind binary model \cite{Eichler:1993}
(Fig.~\ref{fig:147}). This model is supported further by the dramatic
variations of the synchrotron radio emission in the 7.9-year WR+O
binary system WR~140, that are clearly modulated by the binary orbit
(c.f. Fig.~\ref{fig:wr140data}).  A CWB origin for the synchrotron
emission in WR stars was first proposed by
\cite{vanderhucht:1992}. This has now been substantiated with over
90\% of WR stars that exhibit non-thermal emission characteristics
being found to be either binary of having a ``nearby'' massive stellar
companion resulting in a WCR \cite{Dougherty:2000b}.

Colliding-wind binary systems present an important laboratory for
investigating the underlying physics of particle acceleration. In
addition to the excellent modelling constraints provided from high
resolution VLBI imaging, radiometry and high energy observations, CWBs
provide access to higher mass, radiation and magnetic field densities
than found in supernova remnants, which have widely been used for this
work.

\begin{figure*}
\includegraphics[bb=570 35 572 771, angle=-90,width=0.5\textwidth]{wr147_eta0.02_inc30.ps}
\includegraphics[width=0.5\textwidth]{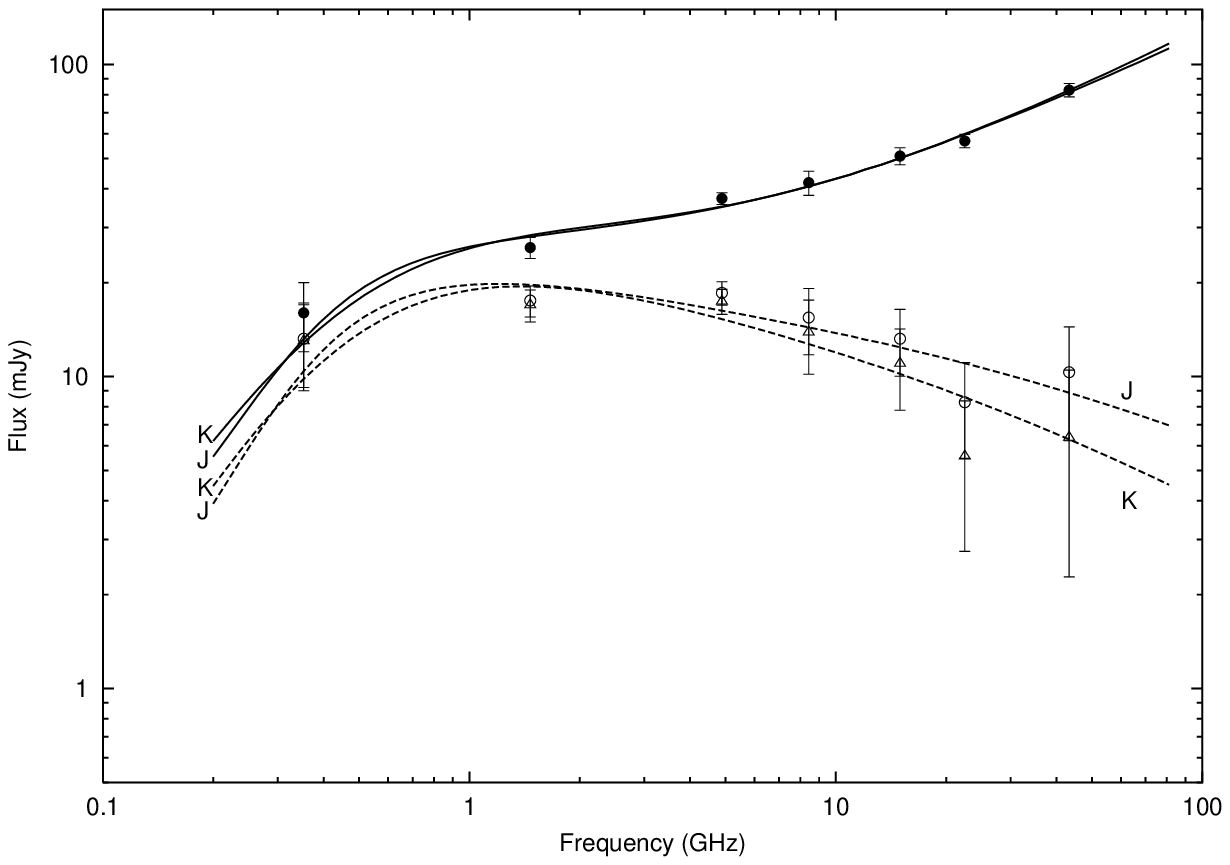}
\caption[]{An example of the density distribution in a model of the
WR147 colliding-wind binary system. The WR star is at (0,0) and the
O-star companion is at ($7.2\times10^{15}$~cm, 0). The WCR bounded by
two standing shocks on either side of the ram pressure balance surface
of the two stellar winds is clearly visible. Right: Radio fluxes from
WR147 with model fits to both the total (solid symbols) and the
non-thermal emission component (open symbols) based on the
2D-hydrodynamic models of the WCR and stellar winds shown on the left
(from \cite{Pittard:2006a}).}
\label{fig:model}
\end{figure*}

\section{Modelling of wind-collision regions} 
\begin{figure*}
\includegraphics[width=0.55\textwidth]{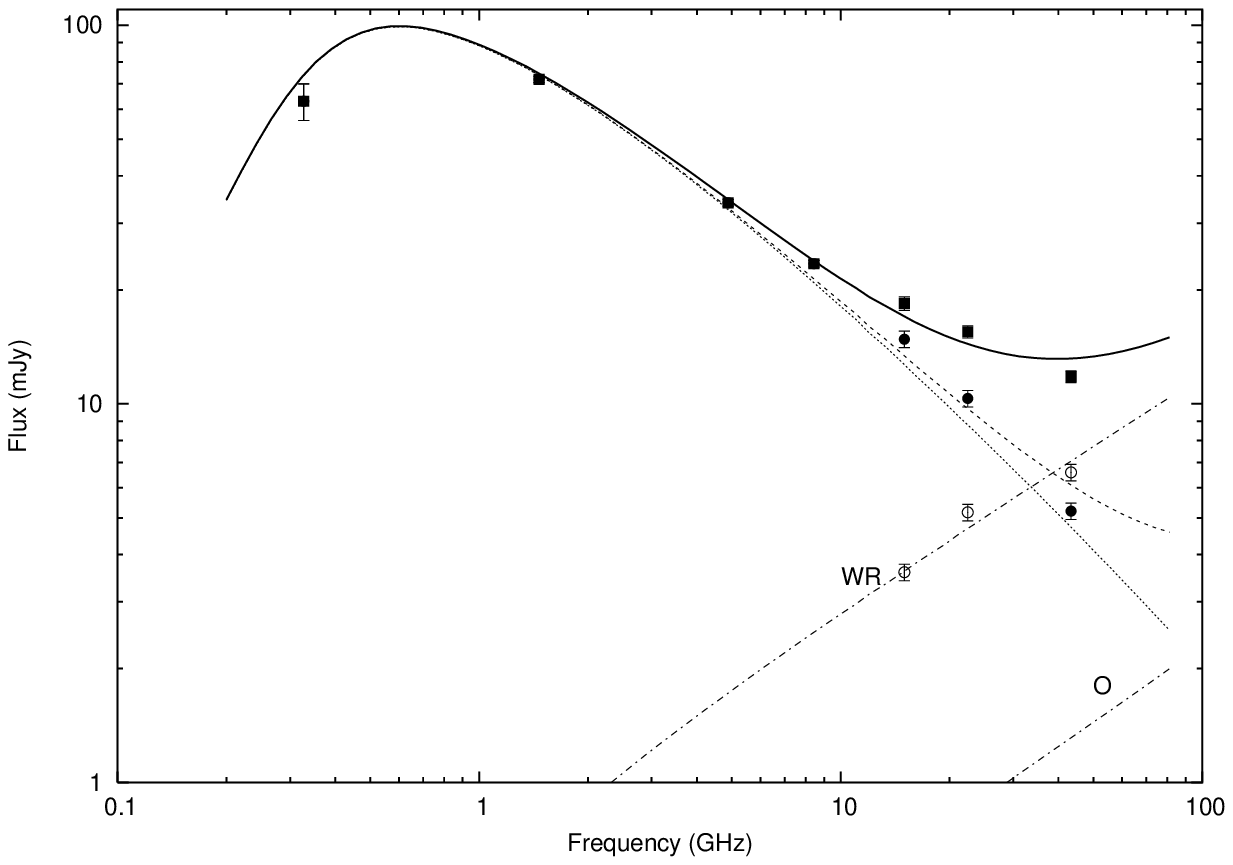}
\includegraphics[bb=38 145 567 640,width=0.45\textwidth]{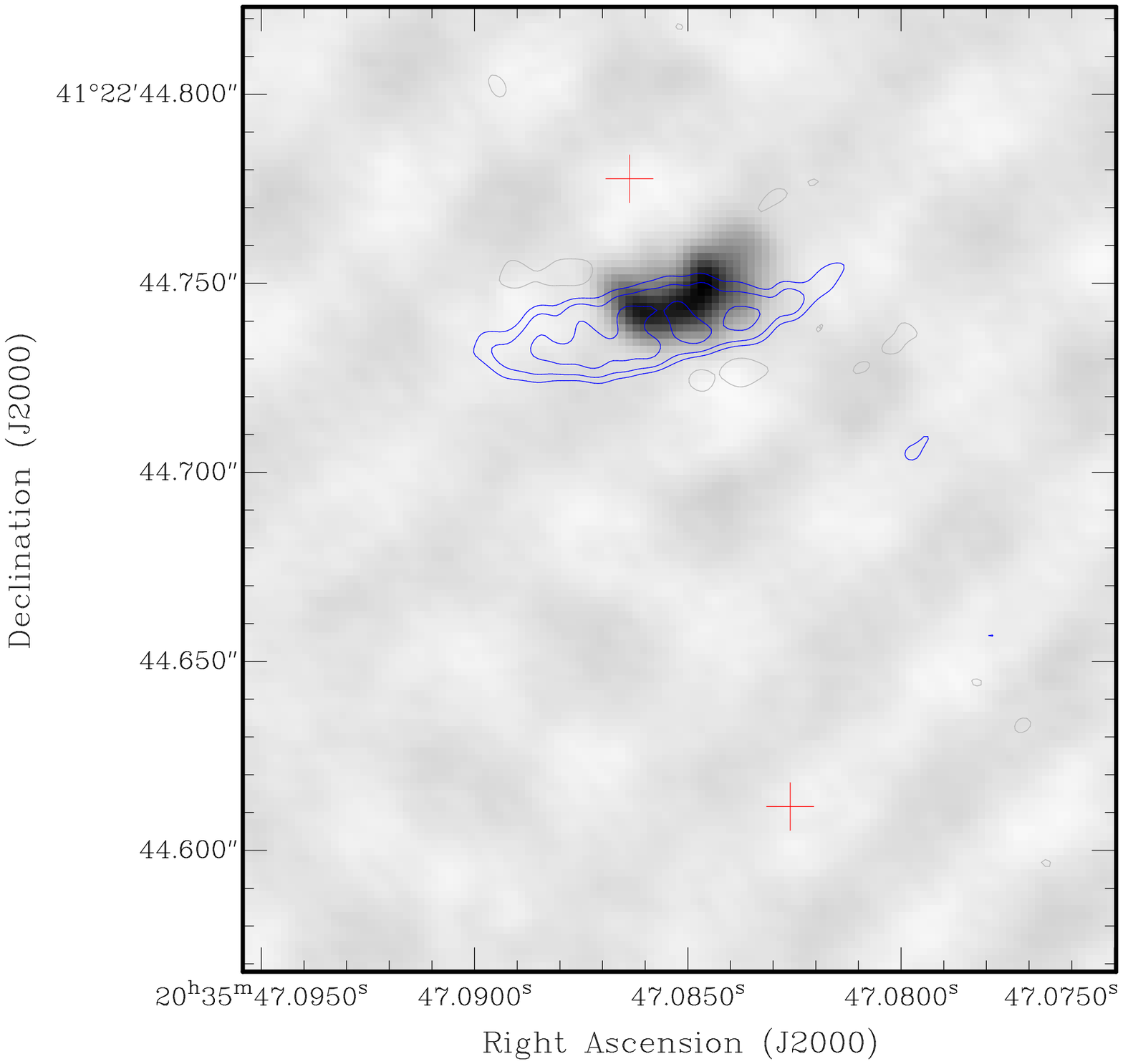}
\caption[]{Left: Best-fit models of the radio emission from the WCR in
WR146, with synchrotron (dotted), thermal (lower dot-dashed) and
synchrotron+thermal (dashed). The thermal emission from the WR star
wind is the higher dot-dashed line.  The solid line is the total
flux. Right: EVN 4.9-GHz observations of the WCR in WR146 (greyscale)
with a simulated EVN observation (contours) based on the model shown
on the left. The crosses mark the relative location of the two stars
(from \cite{O'Connor:2005}).}
\label{fig:146}
\end{figure*}

Models of the radio emission from these systems have been based
largely on highly simplified models, in order to maintain analytic
solutions to the radiative transfer equation. Today, models based on
2-D hydrodynamical models of the density and pressure distributions in
the stellar winds and the WCR are available (Fig.~\ref{fig:model}),
that lead to a description of both the population and the spatial
distribution of the relativistic particles in the WCR
\cite{Dougherty:2003, Pittard:2006a}. In addition to the free-free
opacity of the stellar winds, these models account for changes in the
intrinsic synchrotron luminosity and the free-free and synchrotron
absorption within the WCR. The models assume that the electrons are
shock accelerated by the standing shocks on either side of the contact
discontinuity of the two stellar winds. Other potential acceleration
mechanisms have been considered, such as magnetic reconnection
e.g. \cite{Jardine:1996}, but it has been argued that
although such a mechanism could provide the radio synchrotron
luminosity, reconnection may not provide sufficient
power for the anticipated flux of non-thermal X-ray emission \cite{Pittard:2006b}.  At the
shocks, the energy distribution of the relativistic electrons is
specified by a power law i.e. $n(\gamma)\propto\gamma^{-p}$. As the
electrons advect downstream, the spectrum evolves away from a simple
power law as the electrons cool via Inverse Compton (IC) cooling. The
non-thermal and magnetic energy densities are both normalized to some
fraction of the thermal energy density, with the normalization factors
determined by fitting the radio spectrum.  These models have been
applied with some success to observations of WR147
(Fig.~\ref{fig:147}) and WR146 (Fig.~\ref{fig:146}). The models of
WR147 are not as well constrained by the available data as those of
WR146, since the synchrotron emission does not dominate the total
emission from the system, as in WR146. However, in the latter system
current models fail to accurately reproduce the spectrum at higher
frequencies, and the model WCR emission is more extended than observed
by the EVN \cite{O'Connor:2005}. In spite of these remaining issues,
the value of the modelling constraints posed by both the radiometry
and the VLBI imaging is clear.

\section{WR140 - the particle acceleration laboratory}

In spite of the success of reproducing both the spectra and the
spatial distribution of the radio emission, the models of very wide
systems such as WR146 and WR146 contain a number of ill-constrained,
yet key parameters. These systems do not exhibit radial velocity
variations and so do not have defined orbits. The
inclination of the system to the line of sight is unknown and hence
the geometry of the system, particularly the stellar separation, is
unknown.  Closer systems with periods of a few years present much
better systems for WCR modelling since many orbit parameters are
defined, and they maintain optically thin lines of sight to the WCR.

\begin{figure*}
\begin{center}
\includegraphics[bb=50 70 410 302, width=0.6\textwidth]{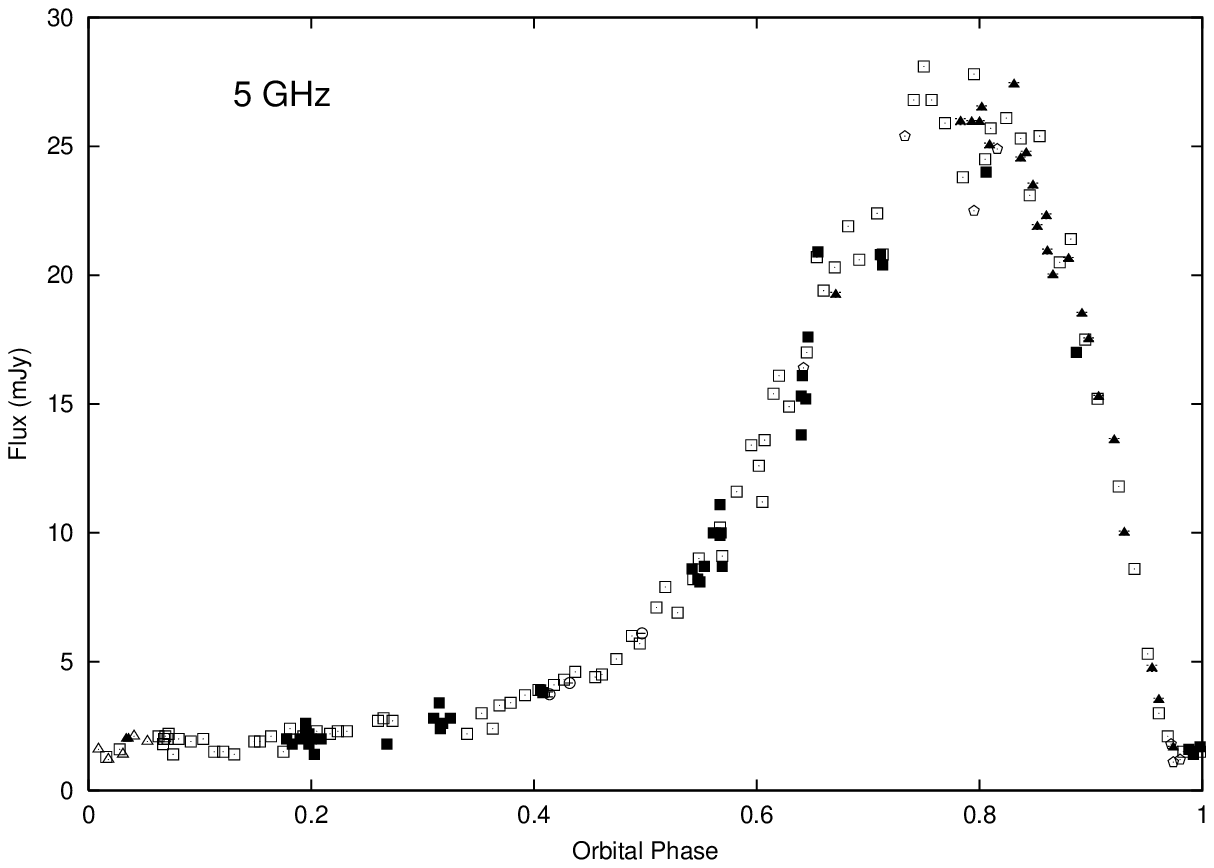}
\includegraphics[bb=0 20 406 550, width=0.37\textwidth]{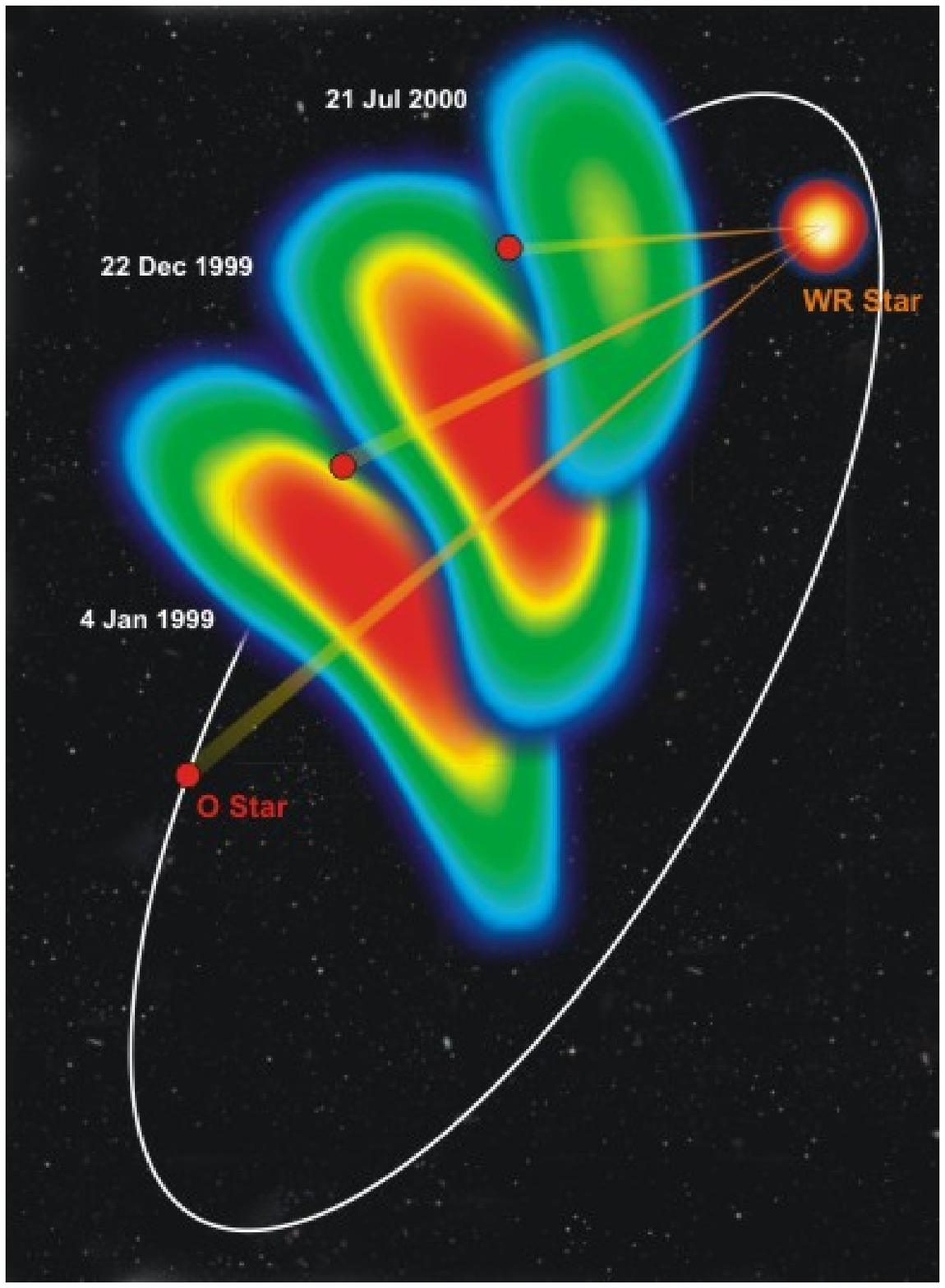}
\end{center}
\caption[]{Left: 5-GHz VLA radiometry of WR\thinspace140 as a function
 of orbital phase for orbit cycles between 1978-1985 (pentagons),
 1985-1993 (squares), 1993-2001 (triangles), and the current cycle
 2000-2007 (circles). Open symbols are from the VLA 
\cite{Dougherty:2005,White:1995} and solid symbols from the WSRT 
\cite{Williams:1990, Williams:1994}. Right: A montage of 8.4-GHz VLBA
 observations of WR\thinspace140 at three orbital phases showing the
 rotation of the WCR as the orbit progresses . The deduced orbit is
 superimposed \cite{Dougherty:2005}.
\label{fig:wr140data}}
\end{figure*}

WR140 is the archetype CWB, with a WC7 star and an O4-5 star in a
highly elliptical orbit ($e \approx 0.88$). It is well known for the
dramatic variations in its emission from near-IR to radio wavelengths
\cite{White:1995,Williams:1990}, and also at X-ray energies during
its 7.9-year orbit \cite{Pollock:2002,Pollock:2005,Zhekov:2000}.  The
WCR in WR140 experiences significant changes as the stellar separation
varies between $\sim 2$~AU at periastron and $\sim 30$~AU at apastron.
The observed radio emission increases by up to two orders of magnitude
between periastron and a frequency-dependent peak between orbital
phases 0.65 and 0.85, followed by a steep decline. This behavior
repeats from one orbit cycle to another, suggesting a particle
acceleration mechanism that is well-controlled by the orbit of the
system (Fig.~\ref{fig:wr140data}).  Recent VLBI observations have
provided key constraints on a number of critical system parameters,
including orbit inclination, distance, and luminosity of the O-star
companion, and make WR140 the ideal system for modelling particle
acceleration in a WCR.

\subsection{Defining the orbit through VLBI observations}

To establish the geometry of the WR140 binary system, particularly the
orbital inclination ($i$) and semi-major axis ($a$), but also the
longitude of the ascending node ($\Omega$), the system must be
resolved into a ``visual'' binary. This was achieved with the
Infrared-Optical Telescope Array (IOTA) interferometer
\cite{Monnier:2004}, and together with the established orbit
parameters \cite{Marchenko:2003}, gives families of possible
solutions for ($i,\Omega,a$). Until further interferometric
observations are available, VLBI observations of the WCR provide the
only means currently to determine uniquely $i$, and hence $\Omega$ and
$a$, from the possible families of IOTA solutions.

A 24-epoch campaign of VLBA observations of WR140 was carried out
between orbital phase 0.7 and 0.9 \cite{Dougherty:2005}. At each
epoch, an arc of emission is observed, resembling the bow-shaped
morphology expected for a WCR. This arc rotates from ``pointing'' NW
to W as the orbit progresses.  The WCR emission is expected to wrap
around the star with the lower wind momentum - the O star.  In this
case, the rotation of the WCR as the orbit progresses implies that the
O star moves from the SE to close to due E of the WR star over the
period of the VLBA observations. Thus, the direction of the orbit is
established.  Secondly, the orbital inclination can be derived from
the change in the orientation of the WCR with orbital phase.  Each
($i,\Omega$) family provides a unique set of position angles for the
projected line-of-centres as a function of orbital phase. This gives
$i=122\deg\pm5\deg$ and $\Omega=353\deg\pm3\deg$, leading to a
semi-major axis of $a=9.0\pm0.5$~mas, and the first full definition of
orbit parameters for {\em any} CWB system.  From optical spectroscopy
$a\,\sin i = 14.10\pm0.54$~AU \cite{Marchenko:2003}, and so the VLBA
orbit parameters require a distance of $1.85\pm0.16$~kpc. This
distance is independent of a calibration of stellar parameters
e.g. absolute magnitude, and implies that the O star is a supergiant.

\subsection{Modelling the radio emission from WR140}
With the geometry and stellar parameters in WR140 defined, models of the
WCR are better constrained than in any other known CWB. In addition,
observations of the thermal X-ray emission that arises from the
shocked plasma in the WCR provide the best constraints on the
mass-loss rates of the two winds as a function of the momentum ratio,
$\eta$, of the two winds. Any clumps in the stellar wind flows are
destroyed by the colliding-wind shocks and the post-shock flow is
smooth. Thus, the thermal X-ray emission provides clumping-free mass-loss
estimates for the two stars.

\begin{figure*}
\includegraphics[width=0.5\textwidth]{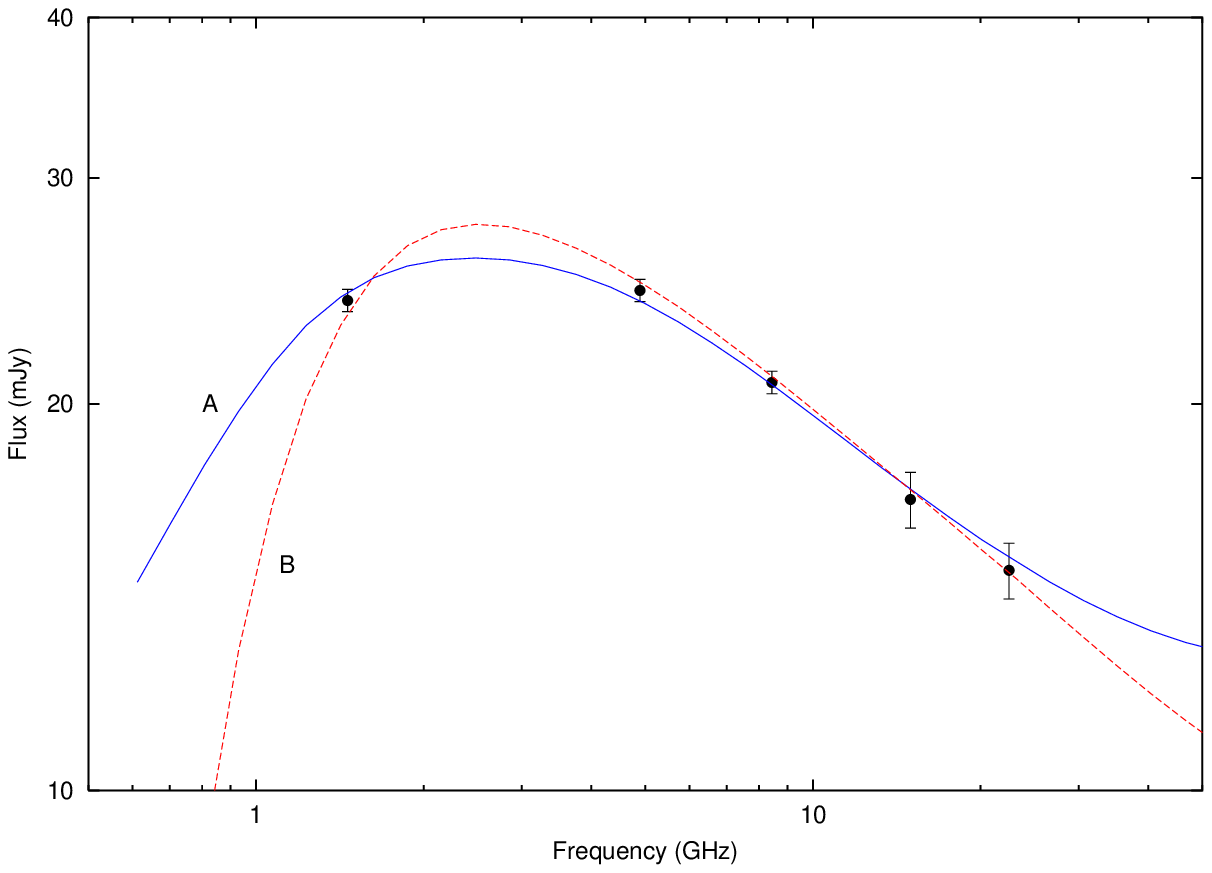}
\includegraphics[width=0.5\textwidth]{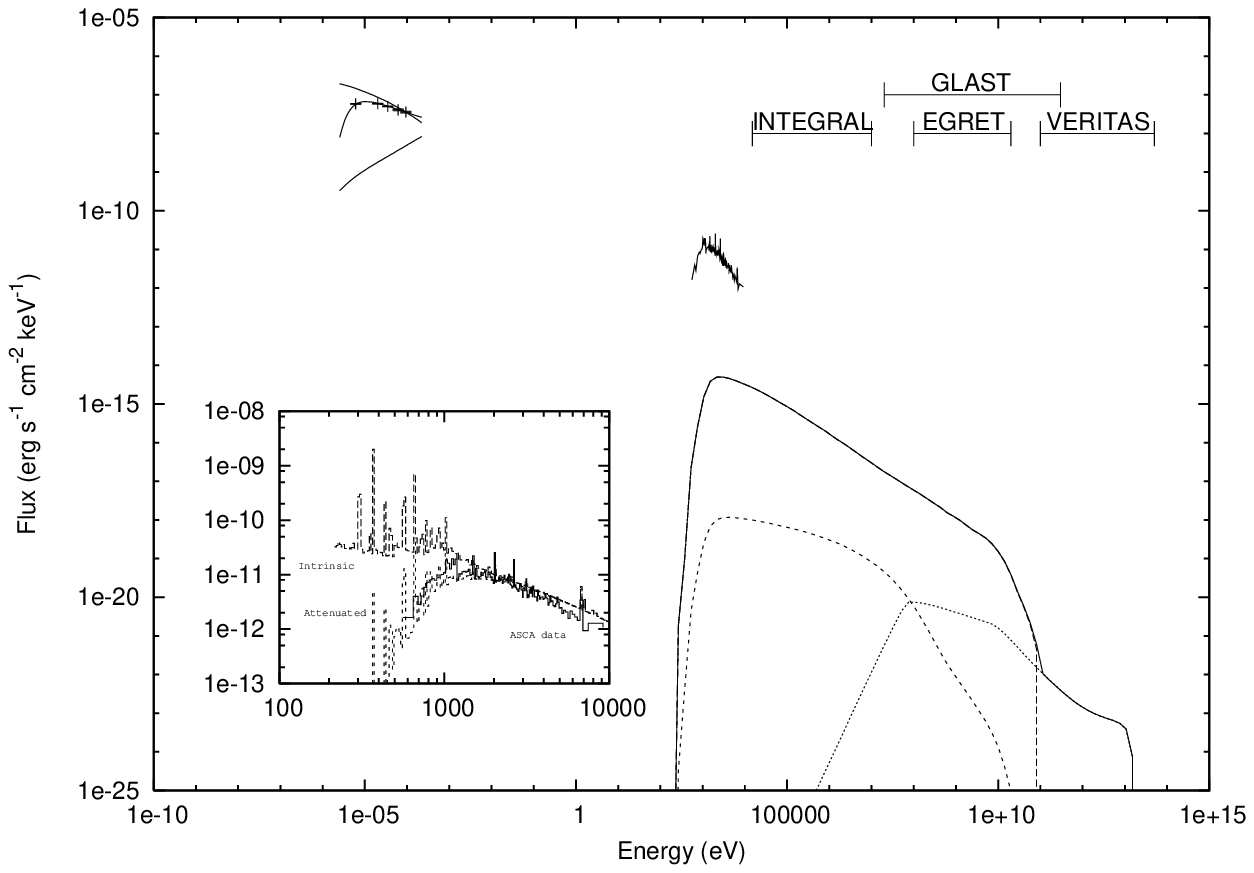}
\caption[]{Left: Model fits to radio data of WR\thinspace140 at
orbital phase 0.837. The observations are the solid circles.  The
wind-momentum ratio of the two winds is 0.22 (model A) and 0.02 (model
B). Equally good fits are possible from both radically different
models. Right: The radio, and non-thermal X-ray and $\gamma$-ray
emission determined from model B, together with the observed radio and
X-ray data (from ASCA). The radio model shown indicates the thermal
free-free flux (displayed below the data points), the {\em intrinsic}
synchrotron flux before free-free absorption (displayed above the data
points), and the observed emission.  The model IC (long dash),
relativistic bremsstrahlung (short dash), and pion decay (dotted)
emission components are shown, along with the total emission
(solid). The energy ranges of a number of satellite missions are
shown. The insert shows detail of model fits to the ASCA thermal X-ray
data that is used to constrain the mass-loss rate and momentum ratio
of the winds (from \cite{Pittard:2006b}). \label{fig:wr140models}}
\end{figure*}

Models of the radio and X-ray emission at orbital phase 0.837 have
been developed \cite{Pittard:2006b}. At this orbital phase,
non-thermal emission dominates the radio spectrum, and so a good
estimate of the characteristics of the relativistic electrons can be
made. Fig.~\ref{fig:wr140models} shows two preferred fits to the
radiometry, with quite different values of relative wind-momentum.
Unfortunately, VLBI arrays do not currently have enough sensitivity to
distinguish between these models through determining the asymptotic
shock opening angle, but the orbit geometry imposed by the multi-epoch
VLBI imaging suggest the model with $\eta\approx0.02$ is most
appropriate.

The models convincingly demonstrate that the turnover at $\sim 3$~GHz
is due to free-free absorption in the winds, since models where the
Razin effect dominates the low-frequency spectrum place an
unacceptably large fraction of the shock energy in non-thermal
electrons. The radio fits also require electron spectra with $p<2$
i.e.  the spectral index of the non-thermal electron energy
distribution is flatter than the canonical strong shock value of $p=2$
expected for diffusive shock acceleration.  A number of mechanisms
could lead to such an index \cite{Pittard:2006b}.

\subsection{Non-thermal high energy emission in WR140}
Definitive evidence for non-thermal X-ray and $\gamma$-ray emission
from CWBs does not yet exist, but many of the unidentified EGRET
sources appear correlated with populations of massive stars
\cite{Romero:1999}.  Notably, WR140 is located in the outskirts of
the positional error box of 3EG J2022+4317 and raises the possibility
of $\gamma$-ray emission from a CWB.

The population of relativistic electrons that give rise to the radio
synchrotron emission are also responsible for high energy IC and
relativistic bremsstrahlung emission. Thus, model fits to the radio
data naturally provide an estimate of the high energy emission from
these processes. The detailed treatment of the spatial dependence of
the non-thermal electron distribution and magnetic field energy
density are expected to produce more robust estimates of the IC
emission than previously available. In particular, the implied flat
electron spectra $p<2$, compared to almost all previous models with
$p=2$, has an important impact on the IC emission spectral index and
the predicted X-ray and $\gamma$-ray flux. In addition to the shock
accelerated electrons, it is anticipated that ions will also be
accelerated to relativistic energies, giving rise to the possibility
of $\gamma$-ray emission from pion decay.

Fig.~\ref{fig:wr140models} shows a predicted high-energy spectrum for
WR140 \cite{Pittard:2006b}. IC emission dominates at energies below
50 GeV while pion decay emission is evident above this energy.  The
nature of the high energy spectrum is strongly dependent on a number
of key properties of the shock acceleration in the WCR. As examples,
the cutoff in the IC and pion decay emission is dependent,
respectively, on the maximum electron and ion energies that can be
attained at the shocks. Also, the luminosity of the IC and pion decay
emission are both dependent on the efficiency of electron and ion
acceleration at the shocks. Indeed, it has already been possible to
rule out several models of the radio data based on the lack of a
detection of WR140 by INTEGRAL. It is clear that observations with
high-energy emission satellites such as GLAST, AGILE, and Suzuka, and
ground-based facilities like VERITAS or MAGIC will provide additional
constraints on the shock acceleration process, particularly the
acceleration efficiency for ions and electrons, and the spectral index
of the non-thermal energy spectrum, $p$. In turn, these will enable
the identification of unique model fits to the radio spectra. This will
be especially important for successful modelling of the spectra of
WR140 throughout its complete orbit cycle.

\section{And what of O-type stars?}
\begin{figure*}
\includegraphics[bb=48 705 473 698, width=0.55\textwidth]{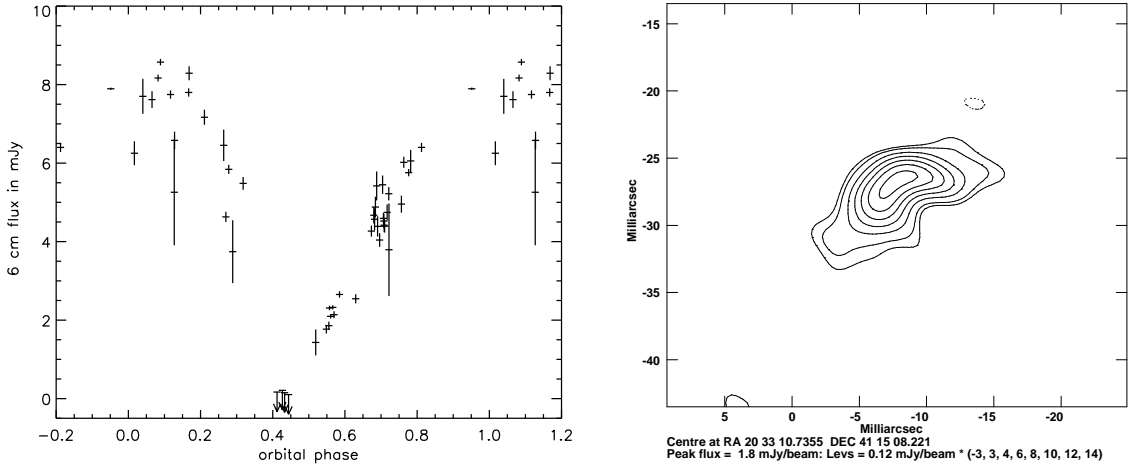}
\includegraphics[bb=50 130 560 682,angle=-90,width=0.46\textwidth]{cygob2x_v2.ps}
\caption[]{Left: 5-GHz flux of Cyg OB2 \#9, modulated by a 2.35-year
period, based on a preliminary reduction of archive VLA data.  Phase
0.0 was arbitrarily set at radio maximum (S. van Loo,
priv. comm.). Right: An 8.4-GHz VLBA observation of Cyg OB2 \#9.  A
bow-shaped WCR is clearly resolved. \label{fig:cygob2}}
\end{figure*}

Now that the source of synchrotron radio emission, and potentially
non-thermal X-ray and $\gamma$-ray emission, in WR stars has been
established as a WCR, it raises the possibility of also explaining
synchrotron emission in O-type stars, the progenitors of WR
stars. Certainly, of the $\sim40\%$ of all radio detected O-stars that
exhibit non-thermal radio emission, 60\% are established binary or
multiple systems \cite{vanLoo:2005}, and a WCR origin for the
emission seems certainly plausible.

Alternative mechanisms have been advanced to explain the non-thermal
emission in O-type stars. Wind-driven instability shocks
e.g. \cite{Chen:1994}, have been demonstrated to be too weak in the
outer regions of the wind to produce the necessary relativistic
electrons if the electrons are trapped by such shocks and cannot be
re-accelerated \cite{vanLoo:2006}. Magnetic confinement of the wind
plasma has been invoked to explain the hard X-ray properties of the
massive stars in the Orion cluster, with the presence of non-thermal
radio emission cited as supporting evidence \cite{Stelzer:2005}.
However, it is certainly clear from high-precision VLBI astrometry
that at least in $\theta^1$ Orionis A the non-thermal radio emission
is {\em not} associated with the massive O-type star, but a nearby
pre-main sequence star \cite{Garrington:2002}. Clearly, VLBI
astrometry is an essential tool in determining the precise location of
the non-thermal emission relative to the stars, most especially when
many of the associations of stars and non-thermal emission were
established initially from relatively low resolution radio imaging
carried out with VLA.

The observational support for the CWB model among O-type stars is
rising.  In the case of Cyg OB2 \#5, a 6.6-day binary O6+O6 star
system, the synchrotron emission originates from a WCR between the
O-star binary and a B-type star $\sim0.9$ arcseconds distant
\cite{Contreras:1997}. However, there remain several O-type stars that
exhibit the characteristics of non-thermal radio emission where a CWB
origin for the emission is challenged by the lack of evidence of a
companion star. In Cyg OB2 \#9, there is no spectroscopic evidence of
a binary system, yet analysis of the radio emission from the system
shows that the emission is clearly modulated with a 2.35-year period
(Fig.~\ref{fig:cygob2}), similar to the behavior observed in WR140. If
Cyg OB2 \#9 is indeed a binary, the secondary must be less luminous
than the primary (otherwise it would have been detected in optical
spectra) and consequently will have a weaker wind. In that case, a
putative WCR should be bow-shaped. VLBA observations of this system
reveal that this is indeed the case (Fig.~\ref{fig:cygob2}), pointing
strongly towards a WCR origin for the emission. The challenge of
finding the companion or companions that give rise to this WCR
remains.

\section{Summary}

High-resolution radio observations have demonstrated that non-thermal
radio emission in WR stars arises where the wind of the WR star
collides with that of a massive stellar companion. There is now
evidence to suggest that non-thermal emission in O-type stars, the
progenitors to WR stars, may also arise from wind collision. In
conjunction with hydrodynamic models of the WCR, radiometry and
high-resolution VLBI imaging are being used to constrain the particle
acceleration process, particularly the electron spectrum and the
efficiency of the acceleration process. These models can be used to
predict the high-energy emission from the WCR. With the advent of high
energy facilities such as GLAST, AGILE, Suzuka, MAGIC, and VERITAS, it
will be possible to place further constraints on the acceleration
process, which will be important for successful models of the dramatic
variations of the emission observed in systems such as WR140.


\end{document}